# One Person, One Bot

Liat Lavi[1]

## Abstract

This short paper puts forward a vision for a new democratic model enabled by the recent technological advances in agentic AI. It therefore opens with drawing a clear and concise picture of the model (section 1), and only later addresses related proposals and research directions (section 2), and concerns regarding feasibility and safety (section 3). It ends with a note on the timeliness of this idea and on optimism (section 4). The model proposed is that of assigning each citizen an AI Agent that would serve as their political delegate, enabling the return to direct democracy. The paper examines this model's relation to existing research, its potential setbacks and feasibility and argues for its further development.

## 1. The Model: One Person, One Bot

Direct democracy is an ideal that is infeasible due to the burden of communication and complexity of legislation. Our reality is one of large societies and complex legal systems. In this reality, it is very hard to imagine how each and every citizen could come to form an opinion on each legislative act, or even just take the time to vote on each matter. Instead, one votes every couple of years (with varied participation rates) in general elections, thereby delegating one's democratic agency to representatives. These representatives are meant to reflect one's political position and social agendas. To say that this system is flawed is a massive understatement.

---

[1] Bezalel Academy of Arts and Design

The current technological developments in AI and more specifically, LLM based agents (the term 'bot' is used in the title mainly for the sake of the linguistic pun) and their role in political debate, is most often referred to as heightening these flaws, raising concerns regarding malicious use, the spread of misinformation through troll bots, the development of new forms of surveillance and oppression and so on (Doshi et al. 2024; Tosi 2025). The model proposed here goes in the opposite, highly optimistic and constructive direction, offering a new paradigm by which AI agents could enable the ideal of direct democracy.

AI agents are algorithmic entities that are especially capable of taking roles and processing large amounts of text. They are thus able to solve some of the central challenges involved in direct democracy - namely the heavy burden of communication involved in it and the complexity of content.

The model I propose is intently simplified - imagine that each and every citizen in a certain society had an AI agent as democratic delegate. This agent would be designed and trained in a manner that reflects its 'owner's' opinions, ethical intuitions, political views, social attitudes and resource-allocation preferences. Its owner would be able to dynamically control its operation, and decide how much freedom to assign it and when to come in and employ its human agency on specific matters.

The 'democratic swarm' of AI agents composed of the bots assigned to each citizen, would process each legislation proposal and vote on it, with each AI agent voting per its owner's preferences.

This process, humanly impossible, would become seamless with the aid of this revolutionary technology, replacing 'one person, one vote' with 'one person, one bot' (henceforth: 'the proposed model').

## 2. Related Ideas and Research Directions

A very similar idea was proposed already by Gudiño in 2018 (also see Lechterman 2025), but had received very little attention in the literature. In a recent paper (Doshi et al. 2024) it is stated that "no literature is yet available on AI replacing politicians", and in a very recent broad analysis of proposals for AI governance (Bullock et al, 2025), the model proposed above is listed as one of several "speculative examples", phrased thus: "Individual direct representation, whereby personalized AGI agents advocate for citizen interests, driving more Pareto-efficient outcomes."

When considering Gudiño and colleagues' (2024) recent version of this idea, it already includes a shift in emphasis, from actual representation to a model of 'digital twins' that are crafted in a generalized manner and enacted in a simulation context.

Bullock et al summarise the proposal thus:

> "Gudiño-Rosero and colleagues (2024) argue that augmented democracy, in this context, "is based on the construction of digital twins, which in the case of this paper, were constructed using a minimalistic representation of each agent (a relatively short vector or demographic characteristics and pairwise preferences)." The idea is to create a simulated version of an individual citizen that can "represent" that person's policy preferences in some arenas of democratic input, as with AI assistants in digital commercial transactions. The authors continue that augmented democracy is: ...different from the creation of AI politicians, as it does not involve the creation of a single AI representative designed to 'listen to everyone,' but an ensemble of AI agents, each controlled by its own human: citizens can create individual profiles that can be personalized according to their own characteristics, preferences and habits, and these autonomous agents can potentially vote on their behalf."

They rightly note that "This work stops far short of the creation of actual "digital twins" for political representation purposes. However, one can imagine personalized LLMs with long-term memory capable of a high-quality simulation (that is, after an extended conversation with the subject and appropriate background information)."

The idea of creating 'digital twins' of objects, human organs and systems, to produce simulations that inform design, decision making and policy emerged from engineering and scientific contexts (Sharma et al. 2022). Today, with the rise of LLM based agents, a path has opened for implementing them in social contexts, by simulating persons and even creating 'digital 'shadow communities' that are meant to augment government by producing simulations that inform policy making (Argota Sánchez-Vaquerizo 2025; Rothe 2024; also see Altera's project Seed 2024).

This line of research is both interesting and important for the proposed model, however due to its limited political claim, it does not invite a discussion of the philosophical facets of the proposed model. As the model proposes a form of heightened human agency through delegating civic duties to AI and thereby awarding greater civic participation, a crucial stepping stone is that of articulating the human-AI agent relationship.

On this point we could draw on existing literature that provides accounts of different status we are to assign to AI agents. Of special import are sources that go beyond the question of metaphysical justification, namely - asking what articulation of AI agents' status is philosophically 'adequate', to also consider the social implications of the structural and semantic choice taken. Such research includes e.g. Joanna Bryson's "Robots Should be Slaves" (2009) and Ethan Mollick's *Co Intelligence* (2024), looking at different positions AI can assume - e.g. as a person, a creative, a coworker, a tutor or a coach.

Particularly relevant to the present model is the notion of AI agents as 'subsidiary agency' (Evans, Robbins and Bryson, 23, p. 10) and of 'surrogate agents' (Evans, Robbins and Bryson, 23, p. 12, Johnson and Powers, 2008). This framework, when applied to the model, highlights that using AI as surrogates is not about creating digital political clones, but rather about 'employing' apt and dedicated political delegates, to work on our behalf and represent the human individual's interests and preferences while employing subject area expertise (similar to hiring a lawyer) and superhuman abilities, such as being able to process large amounts of technical data.

Another important and related field of research is that of Decentralized Autonomous Organizations (DAO), and within it - recent attempts at integrating it with the field of Multi Agent Systems (MAS). As Ao, Liu and Zhang 2025 nicely summarize -

> Decentralized Autonomous Organizations (DAOs) [Wang et al., 2019] are blockchain-based entities governed by encoded rules and protocols, designed to operate without reliance on centralized authorities. DAOs leverage smart contracts—self executing programs on blockchain networks to facilitate transparent, trustless, and autonomous decision-making processes. In the governance framework of a DAO, proposals [Hassan and De Filippi, 2021] serve as the primary mechanism through which members initiate, discuss, and enact changes or decisions. A proposal outlines a specific action or modification, such as funding allocation, protocol adjustments, or governance updates, and is submitted for review by the DAO community. (Ao, Liu and Zhang 2025)

The concept of DAO can be taken in two separate meanings - first, as a general idea of creating agreement based systems that are democratic in nature and operate without a central governance mechanism, or with an automated one. This broad sense is pursued and applied in different arenas, most notably in the context of social organizations and creative endeavors. The second sense of DAO is related to a specific technological development titled Distributed Ledger Technology (DLT), that is most prominently applied in the financial sector through cryptocurrencies and blockchain. The grain, however, of both, has to do with creating 'smart contracts' and facilitating participants' engagement and consensus making mechanisms. The fact that these technologies are applied in the financial sector is in fact promising, as the development of related systems and protocols highlights security and fidelity needs.

However, the concept of DAO is applied to create governance enclaves that are not particularly interested in the existing 'nation-state'. The ideology that initiated this development as well as current applications can even be said to be antagonistic to the idea of the state. Still, if we take the state to be the highest form of 'contract' (Hobbes 1651; Rousseau 1762), and democratic governance as the most crucial form of consensus making, we can easily see the applicable route of DAOs, combined with MAS, as enabling the return to direct democracy proposed by our model.

Also noteworthy are recent developments in the field of LLM based AI, towards creating (non DAO related) mechanisms aimed at consensus making, as in the work of Tessler et al 2024, putting forth a human-oriented LLM based 'Habermas Machine', designed to assist human agents in coming to agreement. This line of research, applied to governance, is also pursued by 'The Computational Democracy Project' (link), developing the 'Polis' platform (also mentioned in Amodei 2024). As the project's site notes:

> Polis is a platform for a conversation, in which participants submit short text statements, or comments, (<140 characters) which are then sent out semi-randomly to other participants to vote on by clicking agree, disagree or pass. Polis allows conversation owners to create conversations which can seamlessly engage (currently) up to hundreds of thousands or (conceivably) millions of participants. (accessed Mar 21, 2025).

These attempts, to date, still rely on human participants' engagement and textual inputs, and thus face capacity and scaling challenges, similar to those that I noted in my opening remarks. They require that the human individual is engaged, reads conversations, writes statements, and so on, and do not (yet) rely on AI surrogate/delegate agents to perform these tasks at scale.

## 3. Setbacks and Concerns

There are numerous concerns that are raised by the proposed model, and I suggest separating them into two categories - concerns having to do with the technology not being good or reliable enough and concerns that will persist even if the technology matures to our satisfaction.

The first class of concerns includes e.g. hallucinations, bias, errors and safety. There is an ongoing disagreement on whether these are 'bugs' or 'features', namely whether these issues can at all be solved or are intrinsic to LLM tech. I leave this discussion aside not because it is irrelevant or unimportant, to the contrary - it is crucial to resolve these problems. Rather, I will not discuss these issues because they receive plenty of attention in the literature, and because the motivation to resolve them could in my view benefit from turning our gaze to better human futures, which may become possible once they are resolved. That is to say that the model proposed is conditioned upon these problems being resolved, but wishes to move further into the future so as to motivate their resolution.

The second class of concerns are issues that remain after the technology has matured, the above mentioned problem resolved, and the model applied. In this speculative scenario, some central issues still need to be addressed. I will relate to these under the headings 'Populism', 'Privacy' and 'Power'.

## Populism

The model proposed heightens the agency of individual citizens, and of 'the public' as an aggregate group of citizens. This raises worries regarding populism, both in its traditional form and in its 'digitally heightened' mode characterizing recent decades. Publics can be manipulated, be fed with disinformation and misinformation, adopt herd behaviours even unknowingly, and are easily shifted towards extreme views. Moreover, when one is asked only for their opinion, without carrying any responsibility for its application and consequences, opinions may easily lean towards caprice and produce irresponsible outcomes.

These worries are justified and do not imply that one in fact does not believe in democracy. The concept of democracy does not boil down to merely following the majority's rule. In its modern form, originating in the French and American revolutions, it included laying ground principles in the form of a constitution, that is hard to change and is set to guarantee that certain values are preserved through shifts in public opinion. For the present model to succeed it would require similarly to set separate mechanisms for making legislative decisions of different orders.

On a more technological level - one would also need to ensure that the surrogate agents are not manipulated directly. This is no easy feat - in order to perform, the agents will need to be informed, and their source of information would mainly be digital media, which due to its distributed structure is particularly susceptible to trends, misinformation, disinformation, extremism and so on. If we are to relocate governance or large sections of it to the digital realm, we are, in principle, enabling large scale and swift manipulations of this sort, thereby heightening these worries.

Since this does not concern merely this model, but rather digital communication at large, this concern is already a well established field of research(see e.g. Williamson, and Prybutok 2024; Yang and Menczer 2025). Existing research highlights that LLM models need to be able to discern which data sources are credible, or minimally, be able to relate to prespecified sources. In the proposed model one may imagine that the data sources are either pluralistic, narrow or personalized, i.e. chosen by the citizen when crafting her personal delegate agent.

## Privacy

Existing voting mechanisms prioritize the individual's right to privacy, and ensure votes are anonymous. The present state of technology awards security breaches that suggest that when the voting mechanism is digitized (also regardless of applying surrogate agents) - privacy and voter anonymity are jeopardized.

There are two replies to this worry. The first reply directs our attention to the most secretive and safeguarded corners of warfare and finance. The most powerful people on earth, in charge of military force, financial systems and so on - are operating in the digital sphere. They are highly concerned about security issues, but they are by and large able to combat these risks with cybersecurity measures. The risk exists, but it is not deterring them from technological advancement. The second reply has to do with how much value we assign to our privacy in general, and our 'political privacy' in particular. The recent decades have shown dramatic changes in how privacy is perceived and valued (see Karniel and Lavie-Dinur, 2017). We have become thoroughly used to being photographed, carry recording devices with us (smartphones), and bringing them into our homes (e.g. Alexa). We broadcast our private lives on social media, and many are also far from shy about their political views on social networks. That is to say that the *desire* to be anonymous in the political context has significantly declined.

Further, the *need* for political anonymity is related to the fear of powerful, antidemocratic regimes. In our present model, where political power is fundamentally distributed, this need may be eradicated.

## Power

Last but far from least - there is the question of power, and here lies a setback that could be detrimental. On the one hand, for the model to be structured, developed and employed would require central organization. On the other hand, if the model proposed is applied, it would imply taking power away from those holding the central organization power, and indeed, from the most powerful. In other words, for the model to actually be applied would require that we elect leaders that are not interested in power, so that they can set up this system and step aside. However, while human history affords revolutions and political coup d'états, it also suggests that those putting their energy into assuming power, also wish to retain it, once achieved. This is the most pressing worry about this model, a worry that merits it the term 'speculative', or even 'paradoxical'.

My only reply to this worry is in the form of a Pascal wager (Pascal 1670) - if the model is not developed, it will most certainly not be applied. If the model, however, *is* developed, there is some chance, slight as it may be, that it will be applied. This is a Pascal wager so long as one sees our present state as leading to hell, and the model proposed as a potential gateway to heaven. I personally think this is the case, and hold this peculiar mix of deep pessimism (about where things are presently heading), and deep optimism (about our ability to change the course of history by applying this model).

This view relies not only on the disappointment from the representational model of democracy to date. It also relies on the recent developments in AI and in the relationship between AI and political power. AI technology in its current form is a new form of power. It is linked to military power, economic power and political power (Crawford, 2021), and currently it is held by people whose sole interest lies in assuming infinite capital, political power, and international dominance. If we are unable to harness this new form of power to the interests of the public and to distribute it effectively, it is sure to make the powerful ever more powerful (Varoufakis 2024).

To complete the picture of the Pascal wager proposed, one also needs to believe that there exists some chance, small as it may be, that this model is applied. Hints in this direction are the fact that large sections of this technology are available as open source, so that one is able to apply them without relying on giant tech companies, and the fact that both in AI and in the realm of DAOs, there exist communities that attempt to develop alternative visions to governance and power distribution. It is a field that has not entirely given up on some of the hopes and promises of Hakim Bey's *T.A.Z* (1991), and on other utopian visions of the digital realm introduced in the previous century by media theorists and Silicon Valley hippies alike. This line of utopian thought may be lost, but not forgotten.

## 4. Hope

> If we want AI to favor democracy and individual rights, we are going to have to fight for that outcome…the vitality of democracy depends on harnessing new technologies to improve democratic institutions, not just responding to risks…. 21st century, AI-enabled polity could be both a stronger protector of individual freedom, and a beacon of hope that helps make liberal democracy the form of government that the whole world wants to adopt. (Anthropic's CEO, Dario Amodei, "Machines of Loving Grace: How AI Could Transform the World for the Better", October 2024)

Since the appearance of ChatGPT we have been subjected to numerous doomsday scenarios and warned about the power of the new phase of AI technology to destroy humanity, or minimally, be exploited by vicious actors to produce immeasurable harms. These scenarios, sounded by both AI protagonists and supporters, should not be disregarded - AI is indeed a new, radical, form of power entering the stage of earth and human civilization.

It is astounding how few positive scenarios are out there, striving towards harnessing this power to advance humanity, and produce alternatives to our current forms of governance and economics. How little intellectual effort is being put in reclaiming the future, and making this form of power not only safe, but beneficial.

The quote with which I opened this section, by Anthropic's CEO, Dario Amodei, cannot be taken at face value. It is true and inspiring, but it is also part of a global campaign in the service of advancing AI tech in the service of its 'owners', the big tech companies holding it. This campaign works to remove regulation, to lower our safeguards, and to create the illusion that the tech giants are working for the benefit of the public. We should not be misled by such statements. It is our job, indeed, our duty, to develop models that, if applied, place humanity on higher grounds, that reduce inequality, that give each of us more, rather than less, agency.

I stated above that the model proposed has been classified as 'speculative' (Bullock et al. 2025), and pointed to its almost paradoxical nature, in the practical sense of how it might come to be employed. Indeed, it may seem very far-fetched. Where does hope lie then? To me, hope lies in the fact that this new technology appeared in parallel to a state of global crisis, a moment of redistribution of power, of war, suffering, shifting ideologies, corruption and also - resistance. On this view, the worse things are, the more hopeful one can be regarding change. The fact that politics has become so disappointing, corruption so deep, economics so unrewarding, the world so chaotic, destruction and war so devastating, while technology is advancing so fast, means that finally - finally! - there is hope.

# References

Amodei, D. "Machines of Loving Grace: How AI Could Transform the World for the Better" (October 2024).

Ao, L., Liu, H., and Zhang, H. "AgentDAO: Synthesis of Proposal Transactions Via Abstract DAO Semantics" (2025). arXiv:2503.10099

Argota Sánchez-Vaquerizo, J., "Urban Digital Twins and metaverses towards city multiplicities: uniting or dividing urban experiences?". *Ethics Inf Technol* 27, 4 (2025).

Bey, H., (Peter Lamborn Wilson), *T.A.Z* (1991)

Bullock, J.B., Hammond, S., and Krier, s. "AGI, Governments, and Free Societies" (2025). arXiv:2503.05710

Doshi, J. et al, "Sleeper Social Bots: a new generation of AI disinformation bots are already a political threat" (2024). https://arxiv.org/abs/2408.12603

Evans, K.D.,Robbins, S.A., Bryson, J.J.,"Do We Collaborate With What We Design?" *Topics in Cognitive Science* (2023) 1-20.

Gudiño, JF, "A bold idea to replace politicians" (2018). https://www.ted.com/talks/cesar_hidalgo_a_bold_idea_to_replace_politicians?language=en

Gudiño JF, Grandi U, Hidalgo, C., "Large language models (LLMs) as agents for augmented democracy." *Phil. Trans. R. Soc.* (2024) A 382: 20240100. https://doi.org/10.1098/rsta.2024.0100

Hobbes, T., *Leviathan* (1651)

Johnson, D. G., Powers, T. M., "Computers as Surrogate Agents", Ch. 13 In *Information Technology and Moral Philosophy*, Van Den Huven, J., Weckert, J., eds. (Cambridge University Press, 2008), 251-269.

Karniel, Y. and Lavie-Dinur, A. *Privacy and Fame* (Lexington, 2017).

Lechterman, T., "Could AI replace politicians? A philosopher maps out three possible futures", *The Conversation*, January 13, 2025, https://theconversation.com/could-ai-replace-politicians-a-philosopher-maps-out-three-possible-futures-246901

Mollick, E., *Co-Intelligence* (Portfolio, 2024).

Pascal, *Pensées* (1670).

Rothe, D., "When the World Is an Object: On the Governmental Promise of a Digital Twin Earth", *International Political Sociology* (2024) 18.

Rousseau, J-J., *The Social Contract* (1762).

Sharma, A., Kosasih, E., Zhang, J., Brintrup, A., Calinescu, A., "Digital Twins: State of the art theory and practice, challenges, and open research questions" *Journal of Industrial Information Integration*, Volume 30 (2022).

Tessler, M. H. et al., "AI can help humans find common ground in democratic deliberation" *Science* 386 (2024). DOI:10.1126/science.adq2852

Tosi, D., Chiappa, M., & Pizzul, D. , "AI Chatbots in Political Campaigns: A Practical Experience in the EU's 2024 Parliament Elections". *Social Science Computer Review* (2025).

Varoufakis, Y,. *Technofeudalism: What Killed Capitalism* (Melville House, 2024)

Williamson, S.M., Prybutok, V., "The Era of Artificial Intelligence Deception: Unraveling the Complexities of False Realities and Emerging Threats of Misinformation", *Information* 15 (2024).

Yang, K-C., Menczer, F. "Accuracy and Political Bias of News Source Credibility Ratings by Large Language Models", arXiv:2304.00228